\documentclass[twocolumn]{article}
\usepackage[a4paper, bindingoffset=0.2in, left=1.5cm, right=1.5cm, top=2cm, bottom=2cm, footskip=.25in]{geometry}
\usepackage{lipsum}% http://ctan.org/pkg/lipsum
\usepackage[utf8]{inputenc}
\usepackage{authblk}
\usepackage{amsmath}
\usepackage{graphicx}
\usepackage{todonotes}
\usepackage{mathtools, cuted}
\usepackage{hyperref}
\usepackage{booktabs}
\usepackage[T1]{fontenc}
\usepackage{lmodern}

\title{\textbf{Numerically Robust Methodology for Fitting Current-Voltage Characteristics of Solar Devices with the Single-Diode Equivalent-Circuit Model}}

\author[1,2,*]{Mario Zinßer}

\affil[1]{Zentrum für Sonnenenergie- und Wasserstoff-Forschung Baden-Württemberg (ZSW), Meitnerstraße 1, 70563 Stuttgart, Germany}
\affil[2]{Light Technology Institute (LTI), Karlsruhe Institute of Technology (KIT), Engesserstraße 13, 76131 Karlsruhe, Germany}
\affil[*]{Correspondence to Mario Zinßer via \underline{mario.zinsser@zsw-bw.de}}

\date{2022 July 6$^\mathrm{th}$}

\newcommand{\kB}{{k_{\mathrm{B}}}}

\newcommand{\Voc}{V_{\mathrm{oc}}}

\newcommand{\Isc}{I_{\mathrm{sc}}}

\newcommand{\Iph}{I_{\mathrm{ph}}}

\newcommand{\Inull}{I_0}

\newcommand{\n}{n_{\mathrm{d}}}

\newcommand{\Rs}{R_{\mathrm{s}}}

\newcommand{\Rsh}{R_{\mathrm{sh}}}

\newcommand{\q}{q_{\mathrm{e}}}

\newcommand{\programLink}{https://github.com/Pixel-95/SolarCell_DiodeModel_Fitting}

\DeclarePairedDelimiter{\ceil}{\lceil}{\rceil}

\begin{document}
\fontfamily{lmss}\selectfont
\twocolumn[{%
\begin{@twocolumnfalse}
\maketitle
\begin{abstract}
For experimental and simulated solar cells and modules discrete current-voltage data sets are measured.
To evaluate the quality of the device, this data needs to be fitted, which is often achieved within the single-diode equivalent-circuit model.
This work offers a numerically robust methodology, which also works for noisy data sets due to its generally formulated initial guess.
A Levenberg–Marquardt algorithm is used afterwards to finalize the fitting parameters.
The source code and an executable version of the methodology are published under \mbox{\url{\programLink}} on GitHub.
This work explains the underlying methodology and basic functionality of the program.\\
\end{abstract}
\end{@twocolumnfalse}
}]

\section{Introduction}
Experimental and simulated solar devices yield a certain current under a given applied voltage and illumination.
Due to the underlying semiconductor physics of p-n junctions, this relation is highly non-linear \cite{shockley1949theory}.
Most of these current-voltage (IV) curves can be described by a simple electronic model with a lumped series and shunt resistance, a diode, and an ideal current source.
The numerical challenge is to find the five fitting parameters for all electric components within the single-diode equivalent-circuit model from a set of experimental data given at $n$ discrete voltages $V_1$, ..., $V_n$ and their corresponding current values $I_1(V_1)$, ..., $I_n(V_n)$.
There exist many algorithms in literature \cite{hegedus2004thin,diantoro2018shockley,toledo2018two,mehta2019accurate} and even their errors are analysed \cite{phang1986review}.
However, the goal of this work is to find a numerically robust method that works also for IV data perturbed by noise effects, such as data outliers, sparse data sets, or higher voltages in modules instead of cells and hence a significantly higher diode ideality factor.

The following section will present the methodology of fitting, while the one after introduces an executable program capable of fitting measured data sets.

\section{Methods}
\begin{figure}[ht]
	\centering
	\includegraphics[width=0.9\linewidth]{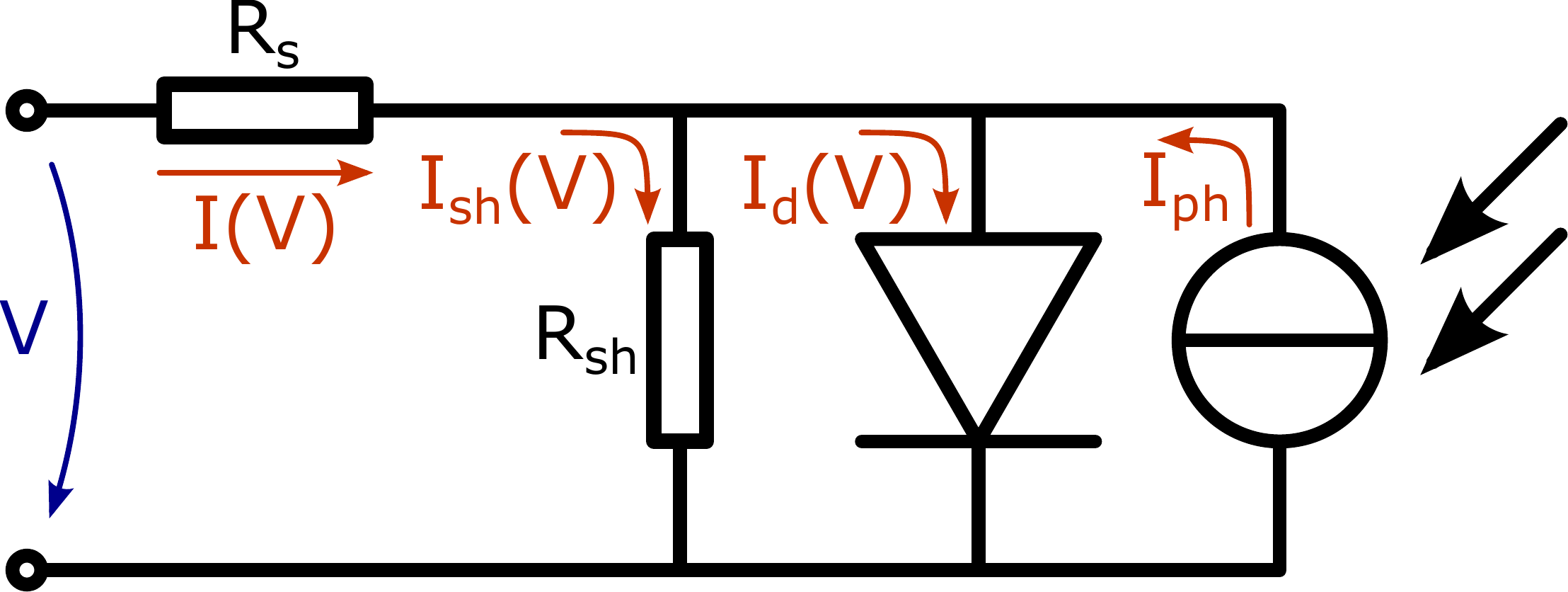}
	\caption{\textbf{Single-diode equivalent-circuit model.} While the photo current represents the generated current, the voltage-dependent shunt and diode currents flow in the opposing direction reducing the produced power.}
	\label{fig:onediodeModel}
\end{figure}
In general, the current-voltage relation\footnote{In this work, all currents are treated as absolute currents $I$ with the SI unit Amperes ($\mathrm A$). However, the same methodology also works for current densities $j$ with the SI unit Ampere per square meter $\left( \frac{\mathrm A}{\mathrm m^2} \right)$.} of regular solar devices at the temperature $T$ as seen in Figure \ref{fig:onediodeModel} can be described by the implicit relation \cite{hegedus2004thin}
\begin{align}
I(V) &= -\Iph + I_\mathrm{sh}(V) + I_\mathrm{d}(V) \notag\\
&=-\Iph + \frac{V - I(V) \cdot \Rs}{\Rsh} \notag\\
&\;\;\;\:\,+ \Inull \cdot \left(\exp\left(\frac{\q \cdot\left(V - I(V) \cdot \Rs\right)}{\n \kB T}\right) - 1\right)\label{diodeEQ}.
\end{align}
Here, $\Iph$ is the generated photo current, $\Rs$ and $\Rsh$ the lumped resistances in series and shunt, $\Inull$ the reverse saturation current, $\n$ the diode ideality factor, $\kB$ the Boltzmann constant, and $\q$ the elementary charge.

By using the Lambert W function \cite{lambert1758observationes} it can be converted into an explicit equation. \cite{jain2004exact,ghani2013numerical,zinsser2022finite}
\begin{align}
	I(V) =\;&\frac{\n \kB T}{\q\Rs} \cdot \mathcal{W}\left(f_{\mathrm{Lam}}\right) + \frac{V - \Rsh \cdot \left(\Iph + \Inull\right)}{\Rs+\Rsh} \label{diodeEQexplicit},
\end{align}
where $\mathcal{W}\left(x\right)$ is the Lambert W function and
\begin{align}
	f_{\mathrm{Lam}} = \;& \frac{\q \Inull \Rsh \Rs}{\n \kB T \left(\Rs+\Rsh\right)} \notag\\
	&\cdot \exp\left({\frac{\q \Rsh \Big(\Rs \left(\Iph + \Inull\right)+V\Big)}{\n \kB T \left(\Rs+\Rsh\right) }}\right).
\end{align}

This section starts with the actual process of fitting and finishes with the calculation of the characteristic data, such as the open circuit voltage from the fitted parameters.

In Table \ref{tab:variables}, all variables used in this work are explained.

\renewcommand{\arraystretch}{1.2}
\begin{table}[ht]
\centering
\begin{tabular}{ l p{6cm} }
\toprule
variable & meaning\\
\midrule
$V$ & voltage \\
$V_i$ & $i$-th voltage of the experimental data set to be fitted \\
$\Voc$ & open circuit voltage \\
$V_\mathrm{MPP}$ & voltage at the maximum power point \\
$I$ & current \\
$I_i$ & $i$-th current of the experimental data set to be fitted \\
$\Iph$ & generated photo current \\
$I_\mathrm{d}$ & current flowing across the diode \\
$I_\mathrm{sh}$ & shunt current \\
$\Iph$ & generated photo current \\
$\Inull$ & reverse saturation current \\
$\Isc$ & short circuit current \\
$I_\mathrm{MPP}$ & current at the maximum power point \\
$P$ & power \\
$P_i$ & $i$-th power of the experimental data set to be fitted (calculated from $V_i$ and $I_i$) \\
$P_\mathrm{MPP}$ & power at the maximum power point \\
$\Rs$ & series resistance \\
$\Rsh$ & shunt resistance \\
$\n$ & diode ideality factor \\
$\mathrm{FF}$ & fill factor \\
$\q$ & elementary charge \\
$\kB$ & Boltzmann constant \\
$\mathcal{W}(x)$ & Lambert W function \cite{lambert1758observationes} \\
$\mathcal{L}_i$ & function value of $\mathcal W(x)$ in the $i$-th iteration of approximation \\
$R^2$ & coefficient of determination \\
$n$ & amount of experimental data points \\
\bottomrule
\end{tabular}
\caption{\textbf{Table of all variables and symbols in this work.} All current values can also be replaced by current density values.}
\label{tab:variables}
\end{table}
\renewcommand{\arraystretch}{1}

\subsection{Fitting Procedure}
The fitting process is divided into two sections.
The first one being the initial guess for the start values of the fitting parameters and the second one being the actual fitting algorithm itself.
Since IV characteristics are highly non-linear, getting a sophisticated initial guess is crucial for the subsequent fitting algorithm and therefore is the most important task.
As already described in the introduction, the goal of the outlined initial guess is not to be as precise as possible, but to be as robust and universal as possible.

\subsubsection{Initial Guess}
The initial guess for the fitting algorithm is obtained by the following procedure.
\begin{enumerate}
    \item As a first step, a cubic Savitsky-Golay filter \cite{savitzky1964smoothing} with a window size of 9 is applied to the experimental data in order to smooth data and not be sensitive to outlier data and noise. Moreover, the current values are eventually multiplied by $-1$ in order to put the data in the appropriate quadrant.
    \item The maximum power point (MPP) is roughly estimated as the discrete data point with the maximum power calculated via $P_i = V_i \cdot I_i$.
    \item The open circuit voltage $\Voc$ is estimated as the linear interpolation of last data point with a negative current and the first data point with a positive current. If there are only data points with negative currents it is calculated via $\Voc = 1.2 \cdot V_\mathrm{MPP}$.
    \item The diode ideality factor is estimated to be $\n = 2 \cdot \frac{\Voc}{\left[\mathrm{Volt}\right]}$.
    \item All data points with a voltage below 20\,\% of $\Voc$ are fitted with a linear regression. The inverse slope is considered as the shunting resistance $\Rsh$ and the y-intercept as the photo current $\Iph$.
    \item The 5 data points with the largest voltage are fitted linearly. The inverse slope of the regression is taken as the initial guess for the series resistance $\Rs$.
    \item Finally, the reverse saturation current is calculated by the diode equation \eqref{diodeEQ} at $V = \Voc$ and hence $I(\Voc) = 0$ via $\Inull = \frac{\Iph - \Voc/\Rsh} {\exp\left(\q \Voc/(\n \kB T)\right) - 1}.$
\end{enumerate}

\subsubsection{Convergent Fitting Method}
Starting from the initial guess described in the past section, the partial derivation with respect to all 5 fitting parameters are derived analytically via the Lambert W function.
Using them, a Levenberg–Marquardt algorithm \cite{levenberg1944method,marquardt1963algorithm} is used in order to perform a regression to all data points.
Since the initial guesses for the photo current and the shunt resistance are typically accurate within a few percent, they are not fitted with the other parameters.
Hence, only $\Inull$, $\n$, and $\Rs$ are fitted in this subprocedure.
At this point, the regression curve usually fits the data points very well.
However, in a second procedure, all 5 parameters are fitted with the Levenberg–Marquardt algorithm to give the program the chance to adapt every parameter at the same time.

\subsection{Calculate Characteristic Data}
Once all fitted diode parameters $\Iph$, $\Inull$, $\n$, $\Rs$, and $\Rsh$ are determined, the solar parameters open circuit voltage $\Voc$, short circuit current $\Isc$, and fill factor $\mathrm{FF}$ need to be calculated.
This section describes their direct determination from the fitting parameters.

\subsubsection{Open Circuit Voltage $\Voc$}
At the open circuit point, the current $I(V)$ vanishes.
Therefore, the open circuit voltage $\Voc$ can be determined by Equation \eqref{diodeEQ} via
\begin{align}
\Voc &= \Rsh (\Iph+\Inull) - \frac{\n \kB T}{\q} \cdot \mathcal{W}\left(f_{\mathrm{Lam}}^{\mathrm{oc}}\right)
\end{align}
with
\begin{align}
f_{\mathrm{Lam}}^{\mathrm{oc}} &= \frac{\q \Inull \Rsh} {\n \kB T} \cdot \exp\left(\frac{\q \Rsh (\Iph+\Inull)} {\n \kB T} \right)\label{eq:Voc}.
\end{align}
If the exponent in Equation \eqref{eq:Voc} is larger than the maximum exponent of double value ($\sim 308$ for IEEE double precision \cite{ieee2019standard}) the Lambert W function $\mathcal{W}\left(f_{\mathrm{Lam}}\right)$ is calculated via the approximation for large numbers given in Appendix \ref{apd:lambert}.

\subsubsection{Short Circuit Current $\Isc$}
The short circuit point is defined as the state without an externally applied voltage and therefore shorted contacts.
Hence, $V=0$ is be plugged into Equation \eqref{diodeEQexplicit} and can be solved via the Lambert W function yielding
\begin{align}
\Isc &= \frac{\n \kB T}{\q \Rs} \cdot \mathcal{W}\left(f_{\mathrm{Lam}}^{\mathrm{sc}}\right) - \frac{\Rsh (\Iph + \Inull)}{\Rsh + \Rs}
\end{align}
with
\begin{align}
f_{\mathrm{Lam}}^{\mathrm{sc}} &= \frac{\q \Inull \Rsh \Rs}{\n \kB T (\Rsh+\Rs)} \cdot \exp\left( \frac{\q \Rsh \Rs (\Iph + \Inull)}{\n \kB T (\Rsh+\Rs)} \right).
\end{align}

\subsubsection{Maximum Power Point MPP}
In order to obtain the power $P$ from a current-voltage characteristic, the current $I$ is multiplied by the voltage $V$.
The maximum power point (MPP) is defined as the point in the IV curve with the highest produced power.
To calculate the MPP, Equation \eqref{diodeEQexplicit} is multiplied by $V$ and afterwards the maximum is determined via 
\begin{align}
\frac{\mathrm dP(V)}{\mathrm dV} = \frac{\mathrm d(I(V) \cdot V)}{\mathrm dV} = 0
\end{align}
by using the Newton-Raphson method.
This yields the MPP voltage $V_\mathrm{MPP}$, resulting in the MPP current $I_\mathrm{MPP} = I(V_\mathrm{MPP})$ via Equation \eqref{diodeEQexplicit} and MPP power $P_\mathrm{MPP} = P(V_\mathrm{MPP})$.

\subsubsection{Fill Factor FF}
Based on the calculation of the MPP, the fill factor $\mathrm{FF}$ is the area ratio of the two rectangles drawn from the origin towards the MPP and from the origin towards the axis intersects.
It is therefore given as
\begin{align}
\mathrm{FF} &= \frac{V_\mathrm{MPP}I_\mathrm{MPP}}{\Voc\Isc}.
\end{align}

\subsubsection{Coefficient of Determination $R^2$}
To be able to tell how good the fit matches the experimental data, a gauge is introduced.
This happens to be the coefficient of determination $R^2$ and is defined as
\begin{align}
R^2 &= 1 - \frac{\sum_i^n \Big(I_i(V_i) - I(V_i)\Big)^2}{\sum_i^n \Big(I_i(V_i) - I_\mathrm{mean}\Big)^2},
\end{align}
where
\begin{align}
I_\mathrm{mean} &= \frac 1n \sum_i^n I_i(V_i).
\end{align}
It typically ranges from 0 (fit always predicts $I_\mathrm{mean}$) to 1 (fit perfectly matches all given data points).
A negative $R^2$ means that the model is even worse than the mean value.

\section{Program}
\begin{figure}[ht]
	\centering
	\includegraphics[width=0.5\linewidth]{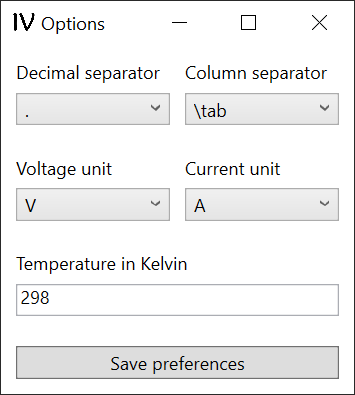}
	\caption{\textbf{Screenshot of the preferences section.} Besides the format symbols and the temperature, the units of the voltage and the current or current density can be selected.}
	\label{fig:options}
\end{figure}

The executable program of the described algorithm is adaptable to certain circumstances as seen in Figure \ref{fig:options}.
Both formatting preferences and units of the voltage and the current (density) can be selected.
Moreover, a certain temperature is required.

\begin{figure*}[ht]
	\centering
	\includegraphics[width=\linewidth]{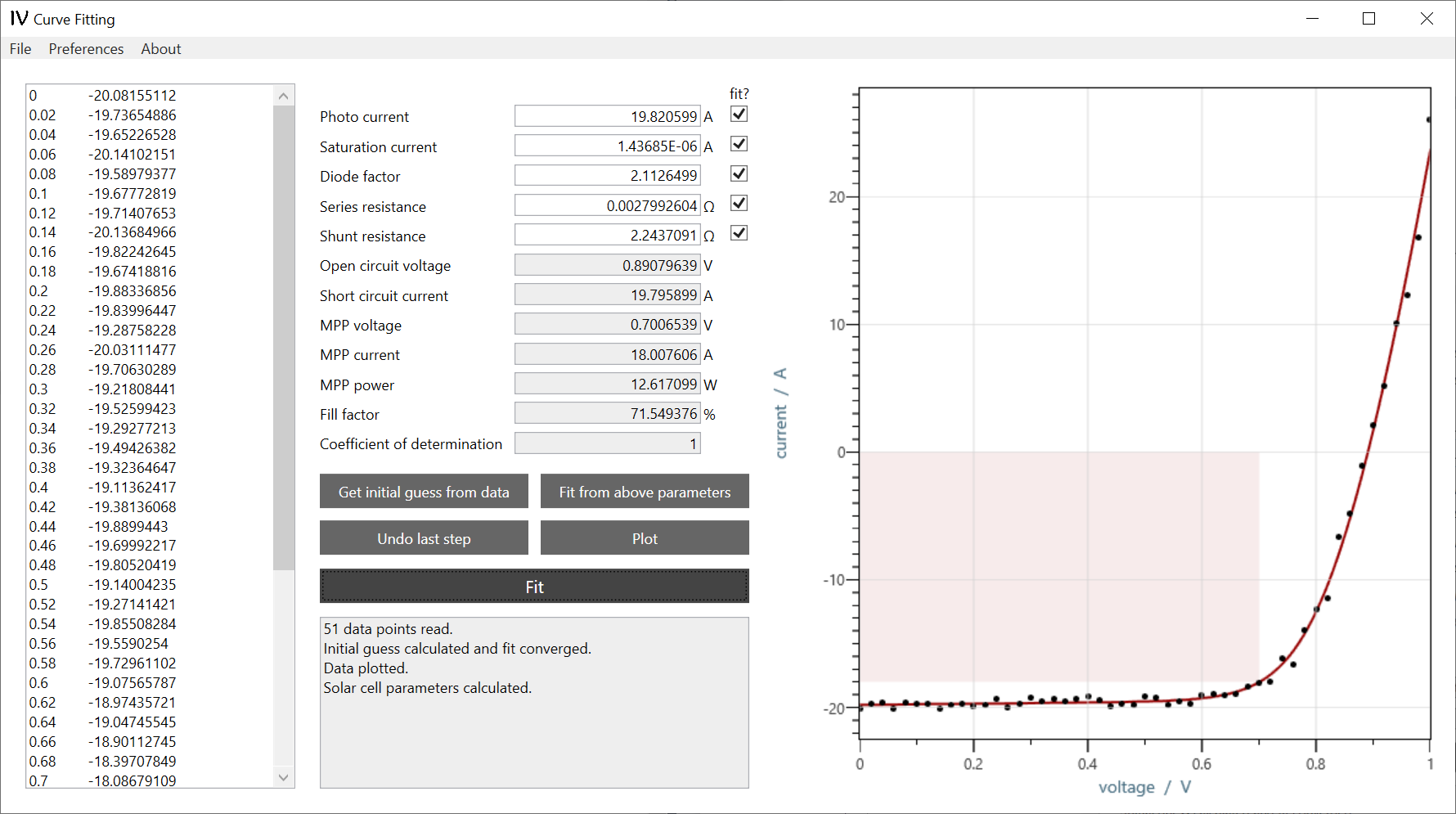}
	\caption{\textbf{Screenshot of the main menu of the program.} On the left side, experimental data is inserted. The middle section contains the fitted diode parameters, the resulting solar cell parameters, all buttons and the logging output box. On the right side, a plot with the experimental data (black points), the fitted curve (dark red) and the area covering the MPP power (light red) are shown.}
	\label{fig:program}
\end{figure*}

The main window of the program is shown in Figure \ref{fig:program}.
Within the left side, the experimental current-voltage data can be inserted.

The middle section shows the fitted diode parameters and the calculated characteristic parameters.
Furthermore, there are all buttons to operate the program.
The main button "Fit" simply fits the experimental data on the left with the above described methodology.
There is also the possibility to only get the initial guess from the data.
The button "Fit from above parameters" does not calculate an initial guess but rather takes the already defined diode parameters as starting values.
The check boxes determine which parameters are fitted.

On the right side, the experimental data is plotted as black points and the fitted curve as red line.
Additionally, a red box represents the covered power at the MPP.

The source code and executable file can be found under \url{\programLink} within a GitHub repository.

\section{Conclusion}
This work introduces a methodology of fitting current-voltage characteristics of solar devices with the single-diode equivalent-circuit model.
Due to the general calculation of the initial guess it is applicable even to perturbed data sets of cells and modules.
Afterwards a Levenberg–Marquardt algorithm is applied in order to convergently determine the required fitting parameters.

\section{Acknowledgements}
The author acknowledges fruitful and productive discussions with Tim Helder, Erwin Lotter, and Andreas Bauer.
This work was inspired and supported by the German Federal Ministry for Economic Affairs and Climate Action (BMWK) under the contract number 0324353A (CIGSTheoMax).

\appendix

\section{Calculating the Lambert W function}\label{apd:lambert}
As the Lambert W function $\mathcal{W}\left(x\right)$ \cite{lambert1758observationes} is defined via the inverse function of the transcendental equation
\begin{align}
\mathcal{W}\left(x\right) \cdot e^{\mathcal{W}\left(x\right)} &= x,
\end{align}
it is necessary to calculate it numerically.
An efficient method to do so is Halley's method \cite{alefeld1981convergence} in order to iteratively approximate its value.
The rough approximation
\begin{align}
\mathcal L_0 = \frac34 \log(x + 1)
\end{align}
is used as an initial guess for the function value $\mathcal L$ and by using the first and second derivatives, the iteration procedure
\begin{align}
\mathcal L_{i+1} = \mathcal L_i - \dfrac{\mathcal L_i \cdot e^{\mathcal L_i} - \mathcal L_i}{e^{\mathcal L_i} \cdot (\mathcal L_i + 1) - (\mathcal L_i + 2) \cdot \dfrac{\mathcal L_i e^{\mathcal L_i} - \mathcal L_i}{2 \mathcal L_i + 2}} \label{eq:iteration}
\end{align}
can be derived.
Equation \eqref{eq:iteration} is iteratively executed $\ceil*{\frac13\log_{10}(x)}$ times with a minimum of four iterations.
This ensures a sufficiently precise accuracy for double precision with a 52 bit long mantissa. \cite{ieee2019standard}

For too large values of $x$, the numbers in equation \eqref{eq:iteration} become too large to be handled by a regular double precision.
Therefore, a simple approximation for large input values is used \cite{hoorfar2007approximation}.
By using the auxiliary variable
\begin{align}
w &= \log\left(x\right)
\end{align}
without any physical meaning, the function value can be approximated via
\begin{align}
\mathcal{W}\left(x\right) = w - \log(w) + \frac{\log(w)}w + \mathcal{O} \left( \left( \frac{\log(w)}{w} \right)^2 \right).
\end{align}

\bibliographystyle{naturemag}
\bibliography{references}

\end{document}